\documentclass{ws-procs9x6}

\begin{document}

\title{RECENT ADVANCES IN NRQCD~\footnote{Talk given at 
7th Workshop on Continuous Advances in QCD,
Minneapolis, Minnesota, 11-14 May 2006. MPI preprint number MPP-2006-157}}

\author{PEDRO RUIZ-FEMENIA and ANDRE~HOANG }

\address{Max-Planck-Institut f\"ur Physik (Werner-Heisenberg-Institut),\\
F\"ohringer Ring 6,
80335 M\"uenchen, Germany\\
E-mail: ahoang@mppmu.mpg.de, ruizfeme@mppmu.mpg.de
}

\begin{abstract}
We discuss recent theoretical  
developments concerning the description of the production and decay 
of heavy quarks and colored scalars in the framework of nonrelativistic QCD. 
\end{abstract}


\bodymatter

\section{Threshold Physics at the ILC}

The $e^+e^-$ center-of-mass (c.m.) energy at a linear collider (LC) can
be very precisely monitored, allowing for an accurate exploration
of the threshold regime.
The top-quark mass can be determined from a measurement of $\sigma(e^+e^-\to Z^*,\, \gamma^*\to
t\bar{t})$ line shape at a LC operating at c.m. energies around the $t\bar{t}$ threshold
($\sqrt{q^2}\sim 350$~GeV). The rise of the cross section with increasing c.m. energy
is directly related to the mass of the top quark. Assuming a total integrated luminosity
of 300~fb$^{-1}$, LC simulations of a threshold scan of the top-antitop total cross
section have demonstrated that experimental uncertainties below 100~MeV for the top-quark mass
determination can be obtained~\cite{Martinez,Hoangetal}, even when beam effects, which lead to some
smearing of the effective c.m. energy, are taken into account.
If the normalization of the cross section line shape is well under control,
it is 
possible to determine the strong coupling, the total top quark width and, if the Higgs
boson is light, the top Yukawa coupling. 
In view of the accuracy obtainable at the LC the theoretical uncertainties
for the total cross section
should be lowered to a level of a few percent~\cite{Hoangtalk,Hoanglogs}.
A precise knowledge of the top mass would also improve the analysis of electroweak
precision observables and put indirect constraints on New Physics.
It has been shown~\cite{NP} that an accuracy of 100~MeV on the top quark mass would
allow to perform   
stringent internal consistency checks 
of the SM and of some scenarios of Supersymmetry (SUSY). 

Threshold studies would also be feasible for squarks at a next LC.
Many models
of SUSY predict that, due to large mixing, the lightest squark could correspond
to one of the mass eigenstates of the third generation with $m_{\tilde{q}} < 500$ GeV,
thus allowing for the production of stop pairs at a future $e^+e^-$ LC
operating below 1 TeV. A ``threshold scan'' of the total cross section
line-shape at such a facility
will yield precise measurements of the stop mass, lifetime and couplings~\cite{stoptalk}, 
in close analogy to the
program carried out in threshold studies for the top-antitop 
threshold~\cite{Hoangetal}.

\section{Theoretical Status of $t\bar{t}$ Production}\label{topreview}

Close to threshold the top quark pairs are produced with small velocities $v\ll 1$ in 
the c.m. frame.
Therefore the relevant physical scales governing the top-antitop dynamics are the top 
quark mass $m_t$, the relative three-momentum $\bm{p}\sim m_t\,v$ and the top quark nonrelativistic
kinetic energy $E\sim m_t\,v^2$. Since the ratios of the three scales can arise in
matrix elements, the cross section cannot be calculated using the standard QCD expansion
in the strong coupling $\alpha_s$. 
The best known indication of the latter comes from
the well-known ``Coulomb singularity",  
which shows up as a singular $(\alpha_s/v)^n$ behaviour in the $v\to 0$ limit
of the $t\bar{t}$ production amplitude at the $n$-loop order
in perturbative QCD. 
The proper expansion scheme for the $t\bar{t}$ threshold region is a double expansion
in both $\alpha_s$ and $v$, and one has to use the parametric 
counting $\alpha_s\sim v\ll 1$ 
to identify 
all effects contributing to a certain order of approximation.
In this 
\textit{fixed-order} expansion the leading order (LO) contributions correspond
to terms in the total cross section proportional to $v(\alpha_s/v)^n$, $(n=0,\dots,\infty)$,
next-to-leading (NLO) terms are proportional 
to $v(\alpha_s/v)^n\times [\alpha_s,v]$ and so on.
At LO the total cross section is proportional to the absorptive part of
the Green function of a Schr\"odinger equation containing the static 
QCD-potential~\cite{Fadin}. 
Higher order corrections in this expansion 
are rigorously implemented employing the concept of nonrelativistic effective quantum field theories, first proposed
by Caswell and Lepage~\cite{Caswell}. In this scheme, the original QCD Lagrangian is
reformulated in terms of an effective nonrelativistic Lagrangian called ``Nonrelativistic
Quantum Chromodynamics" (NRQCD) by using the hierarchy
$m_t\gg \bm{p}\gg E$, which
allows to separate
short-distance physics at the ``hard" scale of order the heavy quark mass from
long-distance physics at the nonrelativistic scales $\bm{p}$ and $E$.
The hard-momentum effects are encoded as Wilson coefficients of the operators 
in the effective Lagrangian. Operators with increasing dimension
are introduced in the effective
theory to include the effects of higher orders in the nonrelativistic expansion,
but only a finite number is needed for a given precision. The NRQCD
factorization properties were used in a number
of NNLO calculations~\cite{HoangTeub2,NNLOcalc} of the total
$t\bar{t}$ production cross section.

A common feature of the \textit{fixed-order} calculations
is that the running from the hard scale $m$ down to the nonrelativistic scales
was not taken into account. At NNLO NRQCD matrix elements and Wilson coefficients 
involve logarithms of ratios of the hard scale and the nonrelativistic scales, 
which in the case of top quark pair production close to threshold can be
sizeable (for example $\alpha_s(m_t)\log[m_t/E]\simeq 0.8$ for $v\simeq 0.15$). 
Considering the parametric counting $\alpha_s\log v\sim 1$ introduces a modified
expansion scheme
for the size of the terms contributing to the total cross section, where the
dominant contribution, proportional to 
$(\alpha_s/v)^n\sum_i^{}(\alpha_s\log v)^i$, $(n=0,\dots\infty)$, is called
leading-logarithmic order (LL). 
A number of different versions
of NRQCD~\cite{BBL,LMR}, each of which aiming (in
principle) on 
applications in different physical situations, allow for renormalization group
improved calculations (see also Ref.~\cite{Reviews} for recent reviews on the field). The EFT vNRQCD
(``velocity''NRQCD)~\cite{LMR,ultrasoft} has been designed for
predictions  at the $t\bar t$  threshold. It treats the case  
$m_t\gg \bm{p}\gg E > \Lambda_{\rm QCD}$, i.e. all physical scales
are perturbative, but also has the correlation $E_t=\bm{p}_t^2/m_t$ built
in at the field theoretical level.
QCD effects for the $t\bar{t}$ total cross section up to the NNLL order 
computed in this framework~\cite{Hoanglogs} significantly reduce the
size and scale dependence to yield a 6$\%
$ theoretical uncertainty.

The effective Lagrangian for the scalar version of
vNRQCD, which  describes the nonrelativistic interaction between pairs of colored scalars
and provides the needed ingredients for a summation of QCD effects at NLL order,
has been given recently~\cite{sNRQCD}.

\section{Finite Width and Electroweak Effects}

Up to now, no systematic and complete treatment of electroweak effects 
in the total cross section for top quark pair production has been achieved 
beyond the LO approximation. The large top
width, being of the same order than the nonrelativistic energy,
is essential in the description of the $t\bar{t}$ threshold dynamics. 
It was shown~\cite{Fadin} that in the nonrelativistic limit the top-quark
width can be consistently implemented by the replacement
$E\to E+i\Gamma_t$ in the results for the total cross section 
for stable top quarks. Although this replacement rule can accommodate
some of the NLO and NNLO electroweak corrections, 
a coherent treatment at the conceptual level 
requires the use of an extended NRQCD
effective theory formalism.

In the NRQCD/vNRQCD framework as long as one is not interested in
any differential information of the top decay,
finite lifetime corrections to the
total cross section can be regarded as short-distance information to be
encoded in the Wilson coefficients of the NRQCD Lagrangian and the NRQCD currents.
As the particles involved in the electroweak
corrections can be lighter than the top quark they can lead to nonzero imaginary
parts in the matching conditions. These electroweak absorptive parts render
the NRQCD Lagrangian non-hermitian, but the total cross section can still be
obtained from the imaginary part of the $e^+e^-\to e^+e^-$ forward scattering 
amplitude by virtue of the optical theorem and the unitarity of the underlying
theory, in complete
analogy to the treatment of the inelastic processes in quantum mechanics
where particle decay and absorption are implemented through potentials with 
complex coefficients.
An important feature
of the effective theory treatment of electroweak effects is that
resonant and non-resonant contributions in NRQCD amplitudes can be systematically
separated if the scaling relation $v\sim\alpha_s\sim \alpha^{1/2}$ is used. The
latter is justified because numerically the top width is approximately
equal to the typical top kinetic energy 
$\Gamma_t\sim m_t\alpha \sim E_{\mbox{kin}}\sim m_t\alpha_s^2$.
Including the non-resonant background diagrams leading to the same final 
states as those of top decay in the matching calculations is necessary in order to
maintain gauge invariance. 

In this
approach the NNLL matching conditions accounting for the absorptive parts related
to the $bW^+\bar{b}W^-$ final
state were derived in~\cite{christoph} and shown to amount numerically to several percent.
A very interesting new conceptual aspect of these corrections is that they have UV divergences 
that arise from the high energy behaviour of the $t\bar{t}$ effective theory
phase space 
integration. The phase-space in the full theory is cut-off by the large
top mass, but it extends to infinity in NRQCD where 
we have taken the limit $m_t\to \infty$~\footnote{Instead of using the
optical theorem to obtain the total cross section from the forward scattering
amplitude one can integrate over the phase-space explicitly with cuts and
these UV divergences obviously do not arise}.
The divergences show up only when $\Gamma_t\ne 0$, as a finite width generates
a distribution for the top invariant mass and thus allows for arbitrary
large momenta in the nonrelativistic phase space integration.   
Similar divergences had already been noted in the QCD NNLL
relativistic corrections to the $S$-wave zero distance Green function if the
unstable propagator was used~\cite{Hoanglogs,HoangTeub2} as well as in the leading order 
$P$-wave zero distance Green function which accounts for $t\bar{t}$ production
through the Z-exchange~\cite{Hoanglogs}. 
The NNLL divergences renormalize $(e^+e^-)(e^+e^-)$ operators that contribute to the total cross
section through the imaginary parts of their Wilson coefficients. The running induced
in these Wilson coefficients by the divergences thus represents a NLL effect to the total cross
section~\cite{christoph} but their matching conditions at the hard scale are presently unknown.

Scalar particles are produced at leading order in the nonrelativistic
expansion in a $P$-wave state. In these systems the phase-space divergences
constitute a more severe problem as they show up
already in the leading order Green's function.
A phenomenologically well motivated example is pair production of 
SUSY partners of the quarks 
at threshold, which has been studied at LO in several works~\cite{Bigistops}
using a semi-phenomenological solution in order to deal with
the phase-space UV divergences.

\section{Nonrelativistic Currents with General Quantum Numbers}

There are a number of issues concerning the consistent formulation of
nonrelativistic interpolating currents which describe color singlet
heavy quark-antiquark and squark-antisquark pair production
for general quantum numbers in $n=3-2\epsilon$ dimensions.
These involve in particular the generalization
of spherical harmonics in $n$ spatial dimensions and the role of
evanescent operator structures ({\it i.e.} that vanish as $\epsilon\to 0$)
for the description of the nonrelativistic spin. The latter have been addressed
in detail in a recent work~\cite{NRcurrents} and we briefly comment on them here.

Let us discuss first
the spin singlet currents with arbitrary angular momentum $L$ (${}^{2S+1}L_J={}^1L_L$).
The generic structure of the production currents with total
spin zero is $\psi_{\bm{p}}^\dagger(x)\,\Gamma(\bm{p})\,\tilde \chi_{-\bm{p}}^*(x)$,
where $\Gamma(\bm{p})$ represents an arbitrary tensor depending on the
c.m.\,momentum label $\bm{p}$ and $\tilde \chi_{-\bm{p}}^*=(i\sigma_2)\chi_{-\bm{p}}^*$.
The interpolating currents associated to a definite angular momentum
state $L$ are related to irreducible representations of the tensor
$\Gamma$ with respect to the rotation group SO(n).
The irreducible tensors are up to normalization just the spherical
harmonics  ${Y}_{LM}(n,\bm{p})$ of degree $L$, with
$M=1,\dots,n_L$, that form an orthogonal basis of a $n_L$-dimensional space with 
$n_L=(2L+n-2)\,\frac{\Gamma(n+L-2)}{\Gamma(n-1)\Gamma(L+1)}$. 
A representation in terms of cartesian coordinates
of the spherical
harmonics of degree $L$ is given by the totally symmetric and traceless
tensors with $L$ indices 
$T^{i_1\dots i_L}(\bm{p})$, where the indices $i_1\ldots,i_L$ are cartesian
coordinates:
\begin{eqnarray}
T^{i_1\dots i_L}(\bm{p}) 
 &=& p^{i_1}\ldots p^{i_L}-{\bm{p}^2\over 2L+n-4}\left(
 \delta^{i_1 i_2}p^{i_3}\ldots p^{i_L}+\dots
 \right)+\dots
\;,
\label{Tdef}
\end{eqnarray} 
and which satisfy the eigenvalue equation for the squared angular momentum operator 
$\bm{L}^2\,\, T^{i_1\dots i_L}(\bm{p})  =  L(L+n-2)\,\, T^{i_1\dots i_L}(\bm{p})$.
The currents with angular momenta S, P and D
are {\it e.g.} relevant in the electromagnetic production of
colored scalars from $e^+e^-$ and $\gamma\gamma$ collisions. 
The use of the generalized currents built from 
(\ref{Tdef}) is mandatory to obtain consistent
results in dimensional regularization in accordance with SO(n)
rotational invariance. An instructive example regarding the computation
of the
nonrelativistic three-loop vacuum polarization diagram with two insertions of the Coulomb potential
is explained in Ref.~\cite{NRcurrents}.

The interpolating currents describing the production of a fermion-antifermion pair
in a spin triplet $S=1$ state for arbitrary $L$ (${}^3L_J$) requires
the treatment of Pauli $\bm{\sigma}$-matrices in $n$-dimensions. The 
$\bm{\sigma}$-matrices $\sigma^i\;(i=1,\ldots,n)$ are the generators of SO(n)
rotations for spin 1/2 and satisfy the Euclidean Clifford algebra $\{\sigma^i,\sigma^j\}=2\delta^{ij}$.
As for the case of the $\gamma$-matrices~\cite{DuganGrin} products of
$\bm{\sigma}$-matrices in arbitrary number of dimensions cannot be reduced to a
finite basis, but represent an infinite set of independent structures, which can 
be choosen as the antisymmetrized product of $\bm{\sigma}$ matrices: 
$\sigma^{i_1\cdots i_m} = \sigma^{[i_1}\sigma^{i_2}\cdots \sigma^{i_m]}\,,\;m=0,1,2,\ldots$
For $m\leq 3$ the
$\sigma^{i_1\cdots i_m}$ are related to physical spin operators, with
eigenvalues of $\bm{S}^2$ equal to $(0,n-1,2,n-3)$, which reduce to the known $n=3$ values.
The $m>3$ operators are evanescent for $n\neq 3$ (although their spin eigenvalues are non-zero). 
S-wave currents in an arbitrary spin state thus have the form
$\psi_{\bm{p}}^\dagger(x)\,\sigma^{i_1\cdots i_m}\,\tilde \chi_{-\bm{p}}^*(x)$, and can arise
in important processes. The structure $\sigma^{i_1 i_2 i_3}$ for example arises in 
fermion pair production in $\gamma\gamma$ collisions, while the evanescent operator
$\sigma^{i_1\cdots i_5}$ is present in the $\bar{f}f\to 3\gamma$ annihilation amplitude.
Note that the differences between
the two different singlet ($m=0,3$) 
and triplet ($m=1,2$) currents correspond to evanescent
operators as well. 

It is well known from subleading order computations based on the effective
weak Hamiltonian that one needs to consistently account for the evanescent
operator structures that arise in matrix elements of physical operators when
being dressed with gluons. A
renormalization scheme can be adopted such that a mixing of evanescent
operators into physical ones does not arise~\cite{DuganGrin}.
Moreover it is also
known~\cite{HerrlichNierste} that modifications of the
evanescent operator basis correspond to a change of the
renormalization scheme. While this does not affect physical predictions, it
does affect matrix elements, matching conditions and anomalous dimensions at
nontrivial subleading order.
Thus precise definitions of the schemes being used have to be given to render such
intermediate results useful. 

In the framework of the nonrelativistic EFT these properties still apply.
However, using the velocity power counting in the EFT allows for even
more specific statements. Concerning interactions through potentials,
transitions between the different S-wave currents built from $\sigma^{i_1\cdots i_m}$
cannot occur because the potentials are
SO(n) scalars and the currents are inequivalent irreducible
representations of SO(n). Even for currents with $L\neq 0$ and for the
spin-dependent spin-orbit and tensor potentials (which is all we need
to consider at NNLL order) one can show 
that transitions between currents
containing $\sigma^{i_1\cdots i_m}$ with a different number of indices
cannot occur~\cite{NRcurrents}. The same arguments apply to the
exchange of soft gluons in vNRQCD. Concerning the 
exchange of ultrasoft gluons, transitions between currents
containing $\sigma^{i_1\cdots i_m}$ with a different number of indices
can arise, but only if the interaction is spin-dependent. The dominant
among these interactions corresponds to the operator 
$\psi_{\bm{p}}^\dagger\sigma^{ij} k^j \psi_{\bm{p}} A^i$ and can only
contribute at N${}^4$LL order, which is beyond the present need and
technical capabilities. 

So for the S-wave currents in $n$ dimensions one can employ either one 
of the two spin singlet ($m=0,3$) or triplet currents ($m=1,2$) in the
EFT and the difference corresponds to a change in the renormalization
scheme. This means in particular that as long as the renormalization
process is restricted to time-ordered products of the currents, one
can freely use three-dimensional relations to reduce
the basis of the physical currents. However, once the basis of the
physical currents is fixed, one has to consistently apply the computational
rules in $n$  dimensions. 
Moreover, one can also conclude that currents containing
evanescent $\sigma^{i_1\cdots i_m}$ matrices ($m>3$) can be safely
dropped from the beginning as long as one does not need to account for spin-dependent
ultrasoft gluon interactions. 

Based on the latter considerations it is
straightforward to construct spin-triplet currents with arbitrary
$L$ (${}^3L_J$). They can be obtained~\cite{NRcurrents}
by determining irreducible SO(n) representations from products of the
tensors $T^{i_1\dots i_L}(\bm{p})$ describing angular momentum $L$ and
the spin-triplet $S=1$ currents discussed previously. 
As for the case of the S-wave currents the physical basis for
arbitrary spatial angular momentum is not unique due to the existence
of evanescent operator structures. It is possible~\cite{NRcurrents} to construct
currents with fully symmetric indices equal to the total angular
momentum $J$ by using the two spin triplet operators, $\sigma^i$ and $\sigma^{ij}$.
There are also currents 
having more than $J$ indices,
which transform according to more complicated patterns of SO(n). They 
become equivalent in $n=3$ to the fully symmetric currents~\cite{NRcurrents}
and are also appropriate to describe production of ${}^3L_J$ states.

The NLL anomalous dimensions of the currents
with arbitrary spin and angular momentum
configurations have been calculated in Ref.~\cite{NRcurrents}. Since at LL
order the currents are not renormalized, their NLL order anomalous dimensions 
are independent of the scheme used for the currents or the potentials.
The NLL running found for the currents 
shows a suppression $\propto \, 1/(2L+1)$, which suggests that the 
summation of logarithms of 
$v$ for the production and annihilation rates of high angular momentum
states is less significant.

\section*{Acknowledgments}

I would like to thank the organizers of the workshop and their crew for the 
pleasant atmosphere during the conference.
 
\bibliographystyle{ws-procs9x6}

\begin{thebibliography}{9}


\bibitem{Martinez}
  M.~Martinez and R.~Miquel,
  Eur.\ Phys.\ J.\ C {\bf 27} (2003) 49
  [arXiv:hep-ph/0207315].
 
\bibitem{Hoangetal}
  A.~H.~Hoang~{\it et al.},
  Eur.\ Phys.\ J.\ directC {\bf 2} (2000) 1.
  
\bibitem{Hoangtalk}
  A.~H.~Hoang, talk at the \textit{International Linear Collider Workshop}, Stanford,
  Califorina, USA, March 2005.

\bibitem{Hoanglogs}
  A.~H.~Hoang, A.~V.~Manohar, I.~W.~Stewart and T.~Teubner,
  Phys.\ Rev.\ Lett.\  {\bf 86} (2001) 1951; 
   A.~H.~Hoang, A.~V.~Manohar, I.~W.~Stewart and T.~Teubner,
  Phys.\ Rev.\ D {\bf 65} (2002) 014014;
  A.~Pineda and A.~Signer,
  arXiv:hep-ph/0607239.

\bibitem{NP}
  S.~Heinemeyer, S.~Kraml, W.~Porod and G.~Weiglein,
  JHEP {\bf 0309} (2003) 075.

 \bibitem{stoptalk}
  I.~I.~Y.~Bigi, V.~S.~Fadin and V.~A.~Khoze,
  Nucl.\ Phys.\ B {\bf 377} (1992) 461;
  H.~Nowak, talk at the \textit{ECFA Linear Collider Workshop}, Durham,
 UK, August 2004;

\bibitem{Fadin}
  V.~S.~Fadin and V.~A.~Khoze,
  JETP Lett.\  {\bf 46} (1987) 525
  [Pisma Zh.\ Eksp.\ Teor.\ Fiz.\  {\bf 46} (1987) 417];
  V.~S.~Fadin and V.~A.~Khoze,
  Sov.\ J.\ Nucl.\ Phys.\  {\bf 48} (1988) 309
  [Yad.\ Fiz.\  {\bf 48} (1988) 487].

\bibitem{Caswell}
  W.~E.~Caswell and G.~P.~Lepage,
  Phys.\ Lett.\ B {\bf 167} (1986) 437.

\bibitem{HoangTeub2}
  A.~H.~Hoang and T.~Teubner,
  Phys.\ Rev.\ D {\bf 60} (1999) 114027; 
  A.~H.~Hoang and T.~Teubner,
  Phys.\ Rev.\ D {\bf 58} (1998) 114023.

\bibitem{NNLOcalc}
  K.~Melnikov and A.~Yelkhovsky,
  Nucl.\ Phys.\ B {\bf 528} (1998) 59;
  O.~I.~Yakovlev,
  Phys.\ Lett.\ B {\bf 457} (1999) 170;
  T.~Nagano, A.~Ota and Y.~Sumino,
  Phys.\ Rev.\ D {\bf 60} (1999) 114014; 
  A.~A.~Penin and A.~A.~Pivovarov,
  Phys.\ Atom.\ Nucl.\  {\bf 64} (2001) 275
  [Yad.\ Fiz.\  {\bf 64} (2001) 323];
  M.~Beneke, A.~Signer and V.~A.~Smirnov,
  Phys.\ Lett.\ B {\bf 454} (1999) 137.

\bibitem{BBL}
  G.~T.~Bodwin, E.~Braaten and G.~P.~Lepage,
  Phys.\ Rev.\ D {\bf 51}, 1125 (1995)
  [Erratum-ibid.\ D {\bf 55}, 5853 (1997)];
  N.~Brambilla, A.~Pineda, J.~Soto and A.~Vairo,
  Nucl.\ Phys.\ B {\bf 566} (2000) 275;
  S.~Fleming, I.~Z.~Rothstein and A.~K.~Leibovich,
  Phys.\ Rev.\ D {\bf 64}, 036002 (2001).

\bibitem{LMR}
  M.~E.~Luke, A.~V.~Manohar and I.~Z.~Rothstein,
  Phys.\ Rev.\ D {\bf 61}, 074025 (2000).

\bibitem{Reviews}
  N.~Brambilla, A.~Pineda, J.~Soto and A.~Vairo,
  Rev.\ Mod.\ Phys.\  {\bf 77} (2005) 1423;
  A.~H.~Hoang,
  arXiv:hep-ph/0204299.

\bibitem{ultrasoft}
  A.~H.~Hoang and I.~W.~Stewart,
  Phys.\ Rev.\ D {\bf 67}, 114020 (2003).

 \bibitem{sNRQCD}
  A.~H.~Hoang and P.~Ruiz-Femenia,
  Phys.\ Rev.\ D {\bf 73} (2006) 014015.

\bibitem{christoph}
  A.~H.~Hoang and C.~J.~Reisser,
  Phys.\ Rev.\ D {\bf 71} (2005) 074022.

\bibitem{Bigistops}
  I.~I.~Y.~Bigi, V.~S.~Fadin and V.~A.~Khoze,
  Nucl.\ Phys.\ B {\bf 377} (1992) 461;
  N.~Fabiano,
  Eur.\ Phys.\ J.\ C {\bf 19} (2001) 547.
 
\bibitem{NRcurrents}
  A.~H.~Hoang and P.~Ruiz-Femenia,
  arXiv:hep-ph/0609151.

\bibitem{DuganGrin}
  M.~J.~Dugan and B.~Grinstein,
  Phys.\ Lett.\ B {\bf 256}, 239 (1991).

\bibitem{HerrlichNierste}
  S.~Herrlich and U.~Nierste,
  Nucl.\ Phys.\ B {\bf 455}, 39 (1995);
   K.~G.~Chetyrkin, M.~Misiak and M.~Munz,
  Nucl.\ Phys.\ B {\bf 520}, 279 (1998).

\end{thebibliography}

\end{document}